%% file: ms.tex
\documentclass[a4paper, 10pt, twocolumn]{article}

\usepackage[english]{babel}          

\usepackage{graphicx}               
\usepackage[font=footnotesize]{caption}
\usepackage[section]{placeins}
\usepackage{url}                
\usepackage{hyperref}

\usepackage{titlesec}
\usepackage{amsmath}                
\usepackage{amssymb}

\graphicspath{{figures/}}

\begin{document}

\date{\small\today}                                        
\title{
	\vspace{-20mm}\textbf{	\large 
	A novel Three-step Network-based Ecosystem Modelling Framework}
	}
\author{
	Laurin Steidle\(^1\),  Inga Hense\\
	\emph{
		\small
		\(^1\)laurin.steidle@uni-hamburg.de} \\
	\emph{
		\small Universitaet Hamburg, Grosse Elbstrasse 133, Germany}
		}


\twocolumn[
	\begin{@twocolumnfalse}
		\maketitle
		\begin{abstract}
			NEMF is a novel network-based ecosystem modelling framework.
			It is a flexible and easy-to-use tool for modelling ecosystems with
			low- to intermediate complexity. 
			It is designed around the idea of visualizing an ecosystem through a
			network that implicitly defines a set of differential equations.
			These equations are then solved numerically and can also be inverse 
			modelled.
			The framework offers the functionality to handle non-equilibrium, 
			non-linear interactions. 
			\vspace*{10mm}
     \end{abstract}
  \end{@twocolumnfalse}
]

\input{framework_paper_content.tex}

\bibliography{references.bib}
\bibliographystyle{unsrt}

\end{document}

%% file: framework_paper_content.tex
\section*{Introduction}


We present a novel ecosystem modelling framework that is able to represent non-linear interactions and offers inverse modelling capabilities.
To our knowledge it is the first framework that combines these features while still being easy to use.

Research or management questions in ecosystem dynamics are often tackled either with complex, 
specifically tailored ecosystem models or by deducing trends from more general models
that are well understood but do not represent the studied system specifically.

However, both modelling approaches are limited.
The first approach is highly time consuming.
The second inherently contains epistemic errors because the more general model does not necessarily represent the system of interest adequately.

Furthermore, when designing a new ecosystem model one needs to know its
interaction sufficiently well to estimate the parameters (i.e. mortality rates) 
needed to represent them accurately.
However, this is rarely the case. 
Therefore, it is crucial to fit the unknown or unprecisely known parameters to reproduce the dynamics of the observed ecosystem.
This process is generally referred to as inverse modelling or sometimes simply as optimisation or fitting.
In contrast, forward modelling is the process of using a model to deduce predictions of a systems behavior.
Solving the inverse modelling problem is a time consuming challenge 
that comes on top of all the other tasks required when modelling ecosystems.

Frameworks have been developed to simplify both forward and inverse modelling problems.
These provide tools for designing, running, fitting, and analyzing a wide range of models for different applications.
Famous examples in the marine ecosystem community are \textit{FABM} \cite{Bruggeman2014}, \textit{carlibrar}\cite{Oliveros2016}, and \textit{gadget} \cite{Begley2004}
(We provide a detailed overview of the presently available marine ecosystem frameworks \href{https://465b.github.io/frameworks-overview/overview-of-present-marine-ecosystem-modelling-frameworks}{\textit{here}}.\cite{frameworkoverview})
However, they all have major short comings.\\
\begin{enumerate}
  \item Some frameworks use linear differential equations 
  \cite{Melbourne-Thomas2012,Borrett2013,Borrett2014,Trebilco2020,Niquil2012} 
  to describe the interaction between system
  compartments.
  However, relationships within ecosystems are highly nonlinear and thus this 
  approach is too simplistic to describe dynamic biological processes adequately.
  \item The frameworks with non-linear interactions are very complex,
  \cite{Bruggeman2014, Audzijonyte2019}
  due to the functionality they offer and their sheer size.
  This makes them too time consuming to effectively apply them to less complex systems.
  \item The frameworks with non-linear differential equations and manageable 
  complexity are all designed for very-specific narrow applications \cite{Shin2004,Begley2004}. 
  This makes them inapplicable for a all other ecosystem modelling tasks.    
\end{enumerate}



We propose a novel ecosystem modelling framework to overcome these shortcomings.
It is designed to offer an easy to use method for modelling ecosystems with low- to intermediate complexity.
The framework offers the functionality to handle non-equilibrium, non-linear interactions.
Because of the broad research field of ecology and its research challenges,
we designed the framework in python with flexibility in mind.
It provides a set of functions for the typical ecosystem interactions 
but it is also possible for the user to define their own interaction functions without needing to change any of the framework code.
The framework also offers methods that fit any parameter of the model to 
represent the studied system.

For simplicity, the current version of the framework is limited to non-spatially resolved models (box-models).

\section*{Framework Overview}

\begin{figure}
  \centering
  \includegraphics[height=0.75\columnwidth ,width=0.8\columnwidth]{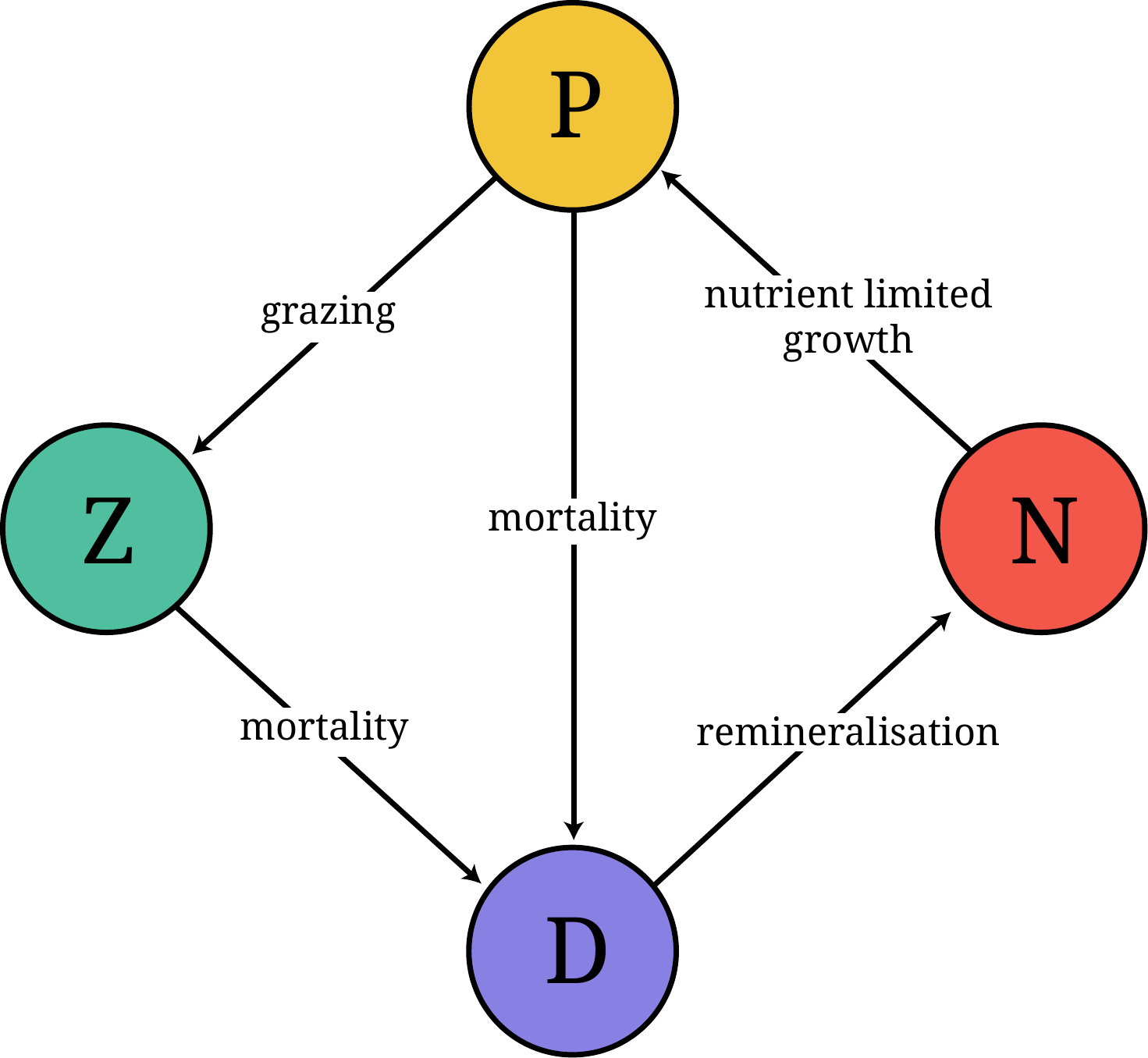}
  \caption{Network visualizing the interactions in an NPDZ model.
           Such a network implicitly defines a a set of differential equations
           \ref{eq:odes}}
  \label{fig:network_diagram}
\end{figure}

Due to the high diversity and complexity in ecosystems,
there are typically no established equations to describe them.
Therefore, conceptual models are a prerequisite for any ecosystem model development.
Thus, typically the first step when modelling an ecosystem is to create a schematic diagram to visualize the
ecosystem compartments and their interactions.
The vertices in such a network represent compartments in the system 
while the links also known as edges represent their interactions.
An exemplary network of an NPZD model which is a simple marine ecosystem model 
is shown in fig. \ref{fig:network_diagram}.
(The so called NPZD model, short for 
nutrient-phytoplankton-zooplankton-detritus model.
See e.g. \cite{Franks1995} for a common representation.)

Through such a type of network one implicitly defines a set of coupled differential equations.
In the case of the presented NPZD model in fig. \ref{fig:network_diagram} they are given by
\begin{equation}
  \begin{aligned}
  \frac{\partial N}{\partial t} & = -J(N)P + \gamma_{m}D \\
  \frac{\partial P}{\partial t} & = (J(N) - \Phi_{P})P - G(\epsilon,g,P)Z \\
  \frac{\partial Z}{\partial t} & = (G(\epsilon,g,P) - \Phi_Z)Z \\
  \frac{\partial D}{\partial t} & = \Phi_{P}P  + \Phi_{Z}Z - \gamma_{m}D  
  \label{eq:odes}
  \end{aligned}
\end{equation}
where the functions $J$,$G$,$\phi$,$\gamma$ 
represent the interactions, namely growth, grazing, mortality, and remineralisation.
The framework provides a standardized way of defining the conceptual model through a simple human-readable configuration file.

A major benefit of this approach is that all interactions can be easily visualized and errors identified.
This is particularly helpful when the systems become more complex and therefore smaller errors might not become immediately obvious.


In the second step, the differential equations are solved numerically.
We offer several numerical solvers for the time integration 
as there exists none that suits all problems.
Additionally, users may also define their time integration methods.

Any parameter used in the forward model can also be inverse modelled.
This may include certain compartment values (i.e. amount of individuals) or interaction parameters (i.e. mortality rates).
To fit the model to a desired behaviour (i.e. measurement data) one only needs to define the upper and lower bound for the particular parameter in addition to providing the data describing the desired behavior.
The inverse problem is then expressed as a non-linear programming problem with fixed constraints.
It is solved by a set of different solvers according to the user's requirement.
The conceptually most simple is the so called gradient-descent approach discussed in the next section.
A description on how the framework is used including an example is found in 
the next section while a more detailed one and further examples can be found in 
the package \href{https://nemf.readthedocs.io}{\textit{Documentation}} \cite{NEMFdoc}.

\newpage
\section*{Description and Example}

This section will use the previously introduced example of an NPZD model to present the usage and technical background of the framework in detail.
%

\subsection*{Forward modelling}

\paragraph{}
The network shown in fig. \ref{fig:network_diagram} contains four vertices representing the compartments. 
Each vertex is associated with a single value of a shared quantity, often referred to as currency.
The choice of currency varies depending on the model. In our example we will use carbon mass in units of kg as our currency.

Each interaction represents a process (e.g. grazing) that is modeled as a flow from one compartment into another.
There exist different types of flows.
For example, mortality based flows depend on one compartment exclusively, while
predator-prey interactions depend on both organism groups.
To account for this, each flow is labelled, representing the specific kind of interaction.

\begin{figure}
  \centering
  \includegraphics[width=0.7\columnwidth]{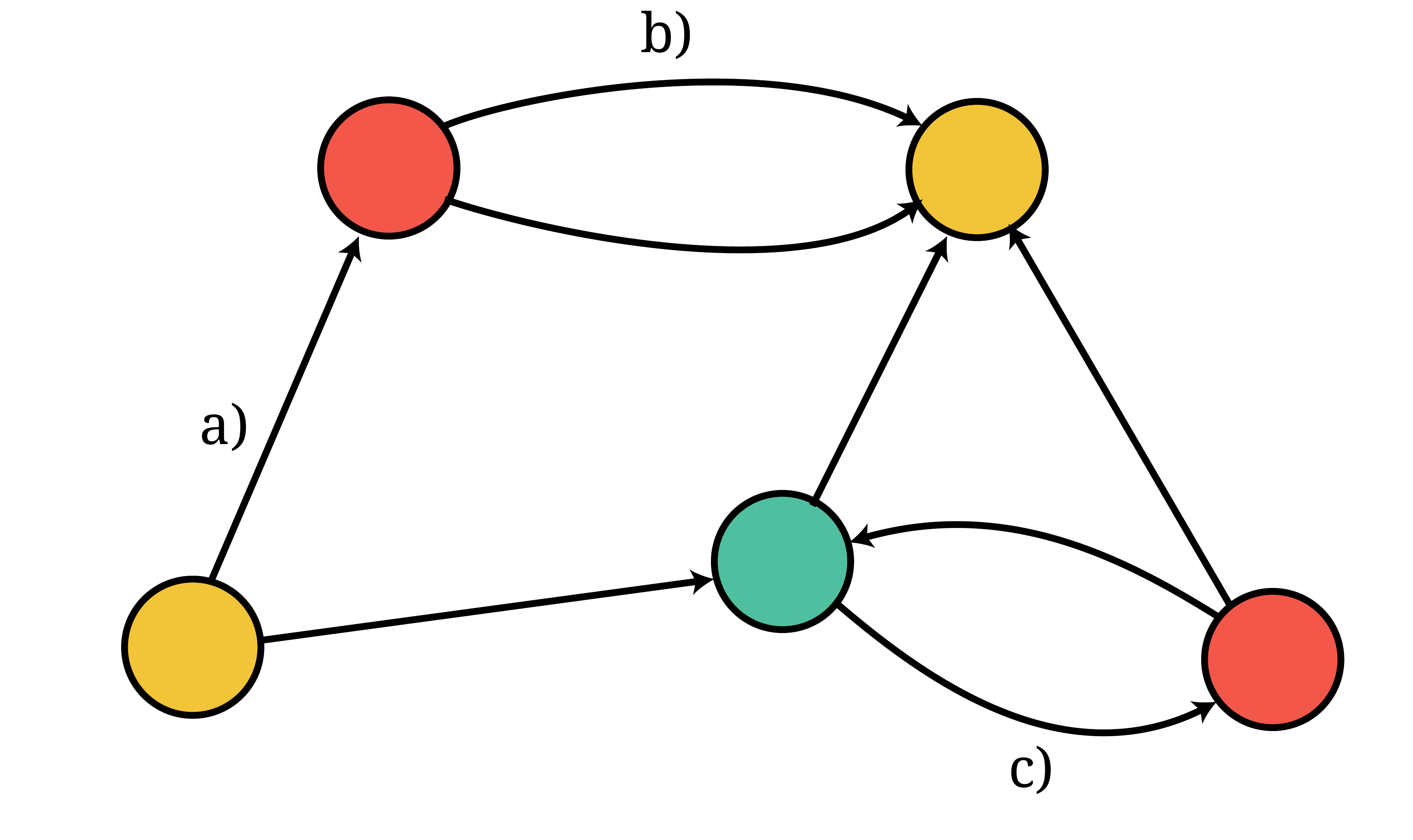}
  \caption{A complete example of a directed multi-graph. a) shows a single directed interaction. b) shows multiple directed interactions. c) shows a bidirectional interaction.}
  \label{fig:graph}
\end{figure}

Furthermore, each vertex can be the origin and the destination of several flows.
The framework also allows that there might be multiple flows between one set of origin destination vertices.
Such an network is referred to as a directed-labelled-multigraph.
A complete example of a directed-labelled-multigraph is shown in fig. \ref{fig:graph}.

We assume that the currency is conserved in all interactions. Hence, all the flow leaving one vertex is always ending up in another.
None is lost in the transfer.
Initially, this might seem like a limiting constrain.
However, if there is a desire to model an ``inefficient'' interaction where parts of the flow are lost, an artificial loss compartment can be introduced.

\paragraph{}
To enable the framework to construct such kind of graph, the user only needs to provide a list of flows, describing origin and destination, its type and parameters used in the description of the flow.
This set of information is expressed in a standardized yaml format \cite{YAML}.
The interaction types must either correspond to the name of an interaction function offered in the frameworks interaction library or to a user written function.

The option to easily add user written alternatives to the present functions is one of the core features of this framework as it offers a great deal of flexibility which is not present in any other similar ecosystem framework.

\begin{figure}
  \includegraphics[width=\columnwidth]{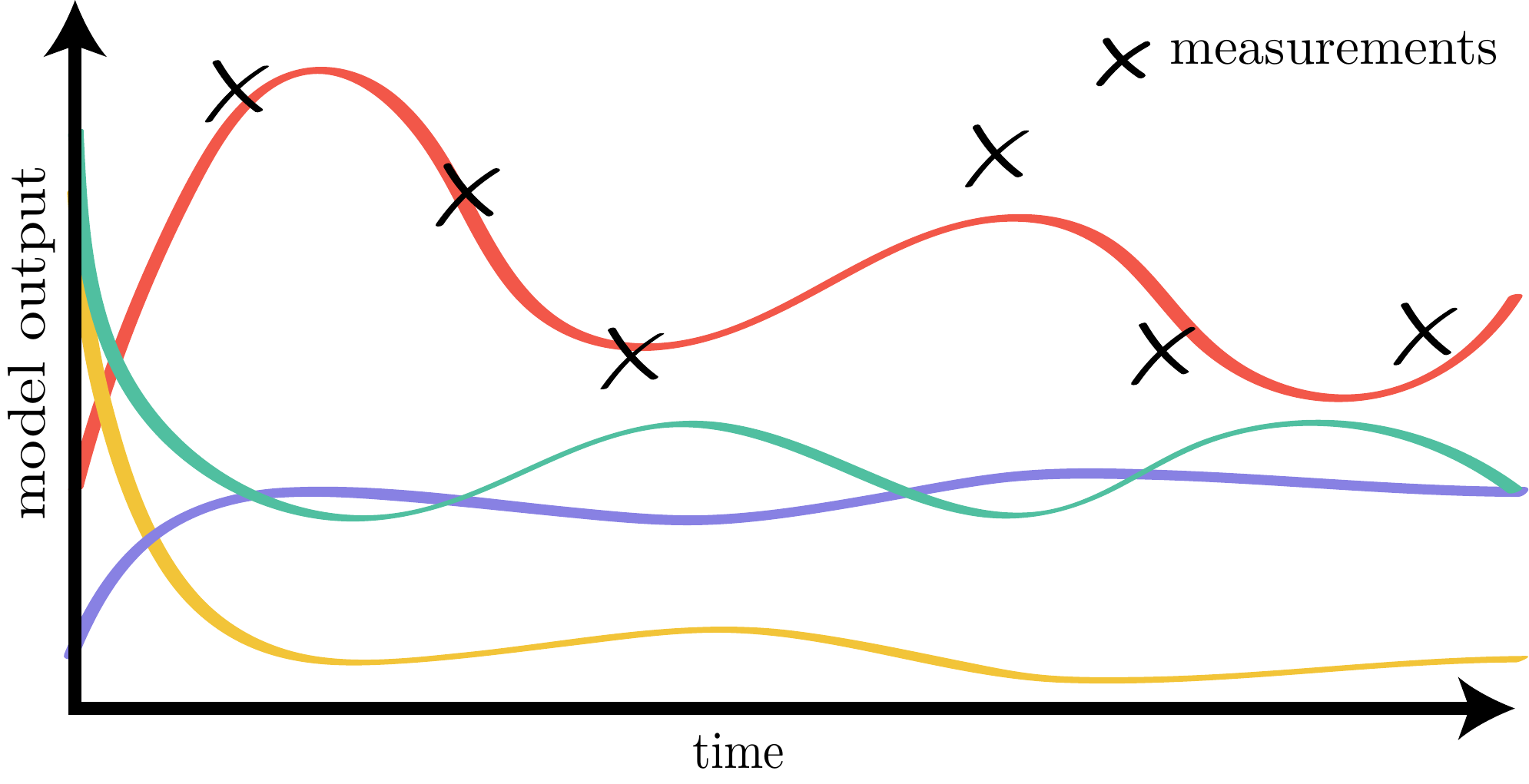}
  \caption{Exemplary solution of the differential equations  shown in equation \ref{eq:odes} as visualized in fig. \ref{fig:graph}. Each color represents one compartments population or mass over time.}
  \label{fig:npzd_time_evo}
\end{figure}

\paragraph{}
A set of differential equations  as shown in eq. \ref{eq:odes} is constructed based on the network structure read from the yaml file.
The type of differential equations that are constructed are non-linear 
first-order coupled ordinary differential equations (ODE).
These equations describe the dynamics of the system deterministically for a given set of initial conditions.
An exemplary numerical solution of eq. \ref{eq:odes} is presented in fig.
\ref{fig:npzd_time_evo}

\paragraph{}
After the model has been designed and its numerical solution is found,
the question arises if this model actually represents the system that it is 
designed to describe.
Only rarely the first guess of the model parameters define the model
in such a way that it mimics the behaviour of the studied system.
A simple example of a system's behaviour might be that the studied system is in a steady-state. 
While the initial guess of the model might perform reasonably well and also reaches a steady-state after a certain build-up phase, it might reach a different one than the studied system.

For less complex models, certain parameters, often referred to as fitting-parameters, can be adjusted by hand until the model mimics the studied system.
However, for more complex models this becomes unfeasible.
Therefore, this so called inverse modelling process needs to be automated.

\paragraph{} 
When fitting the model by hand, its``quality'' is implicitly judged.
To automate this process, an objective quality measure needs to be formalized.
This objective function is often also called loss- or cost-function.
For our framework we define the cost-function $L$ as the euclidean norm,
which is a quadratic measure of distance between the reference data provided for
all compartments \textit{i}, and time \textit{t} and the model output at these 
points.
\begin{equation}
  L(\alpha_j) = ||\bar{x}^{i}_{t}-\tilde{x}^{i}_{t}(\alpha_j)|| = \sqrt{\sum_{i,t} (\bar{x}^{i}_{t}-\tilde{x}^{i}_{t}(\alpha_j))^{2}}
  \label{eq:L2}
\end{equation}
where $\bar{x}$ represents the reference value, and $\tilde{x}$ the model value.
$\tilde{x}(\alpha_{j})$ is a function of the model parameters $\alpha_{j}$ which
are tuned during the optimization process.

\begin{figure}
  \includegraphics[width=\columnwidth]{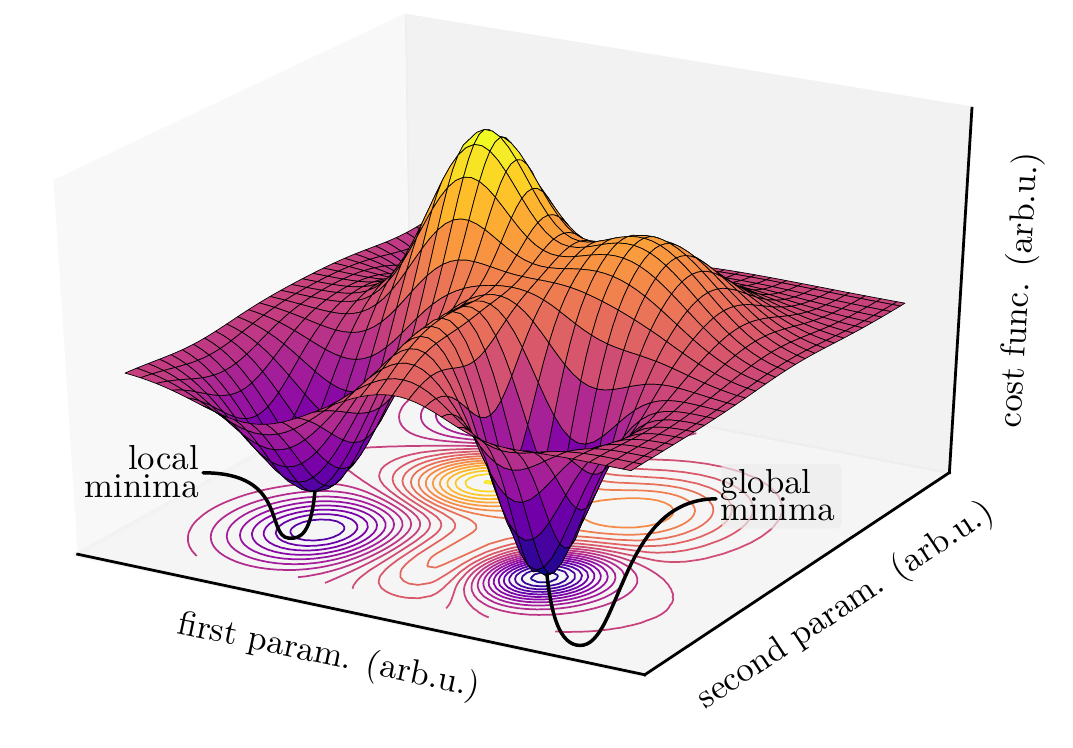}
  \caption{Exemplary cost-function field. Each point on this field represents the measure of quality for a certain configuration of the model. The lower the cost the better the model. Hence, it is the goal of the fitting process to find the model configuration at the global minimum.}
  \label{fig:local_global_minima}
\end{figure}

In a simple example there might only be two free parameters that are varied during the optimisation process.
In that case, each parameter would span one dimension, while the cost at each 
points spans a third dimension as shown in fig \ref{fig:local_global_minima}.

\paragraph{} 
With the help of this cost-function field, which measures the quality of each possible model configuration, a better model can be found through minimization.
Yet the key problem is, that the cost-function is not globally known.
Therefore, we have to explore the "landscape" like a hiker in foggy mountains.
Each individual model configuration for a set of parameters has to be evaluated to measure its performance. Note, that in general we can choose any amount of free parameters creating a multi-dimensional cost-function field.

This class of problem, where a continuous non-linear cost function is minimized, is known as non-linear programming problem. There are several different methods to tackle this problem.

\paragraph{} 
One of the simplest method is a gradient based approach called "gradient descend".
\cite{Curry1944}
The key idea here is that as long as we always walk downhill we will end up at the bottom of the valley eventually.
To ensure that we always walk downhill, we calculate the gradient at a given point by evaluating models with slightly perturbed parameters.
In other words, we take a step in each direction and choose the direction that goes downhill the steepest.
In its simplest implementation, this concept is described by the following equation
\begin{equation}
  \alpha^{(n+1)}_{j} = \alpha^{(n)}_{j} - \lambda \nabla J(\alpha^{(n)}_{j})
\end{equation}
where $\alpha^{(n)}_{j}$ and $\alpha^{(n+1)}_{j}$ represents the current and the "improved" guess of model parameters. $\lambda$ describes the step size that we take.

Typically one does not end up at the ideal model after the first step.
Hence, this process is iterated several times until either convergence is reached, meaning that the model doesn't improve anymore or a satisfying model is found.

A major problem in these optimization schemes is that generally, one can not 
distinguish if the solution of the optimization process is a local or global 
minimum.
Meaning, that there might be a deeper valley on the other side of the mountain that is unknown to us.
(See fig. \ref{fig:local_global_minima}).

Unfortunately, there exists no guarantee that we find the deepest valley.
The only option we have to overcome this problem is by using different model configurations as a starting point of the optimization process, and test if a better solution can be found.
This is known as a monte-carlo approach.

\paragraph{} 
While the inverse problem is a challenging task from a mathematical perspective, it is automated in our framework.
Nevertheless, the user can select several options for this process and also use self-written optimization routines.
The user only needs to provide two things:
\begin{enumerate}
  \item the data that represents the system that shall be represented
  \item to mark the parameters that are allowed to be varied by providing their upper and lower bounds
\end{enumerate}

\begin{figure}
  \centering
  \includegraphics[width=\columnwidth]{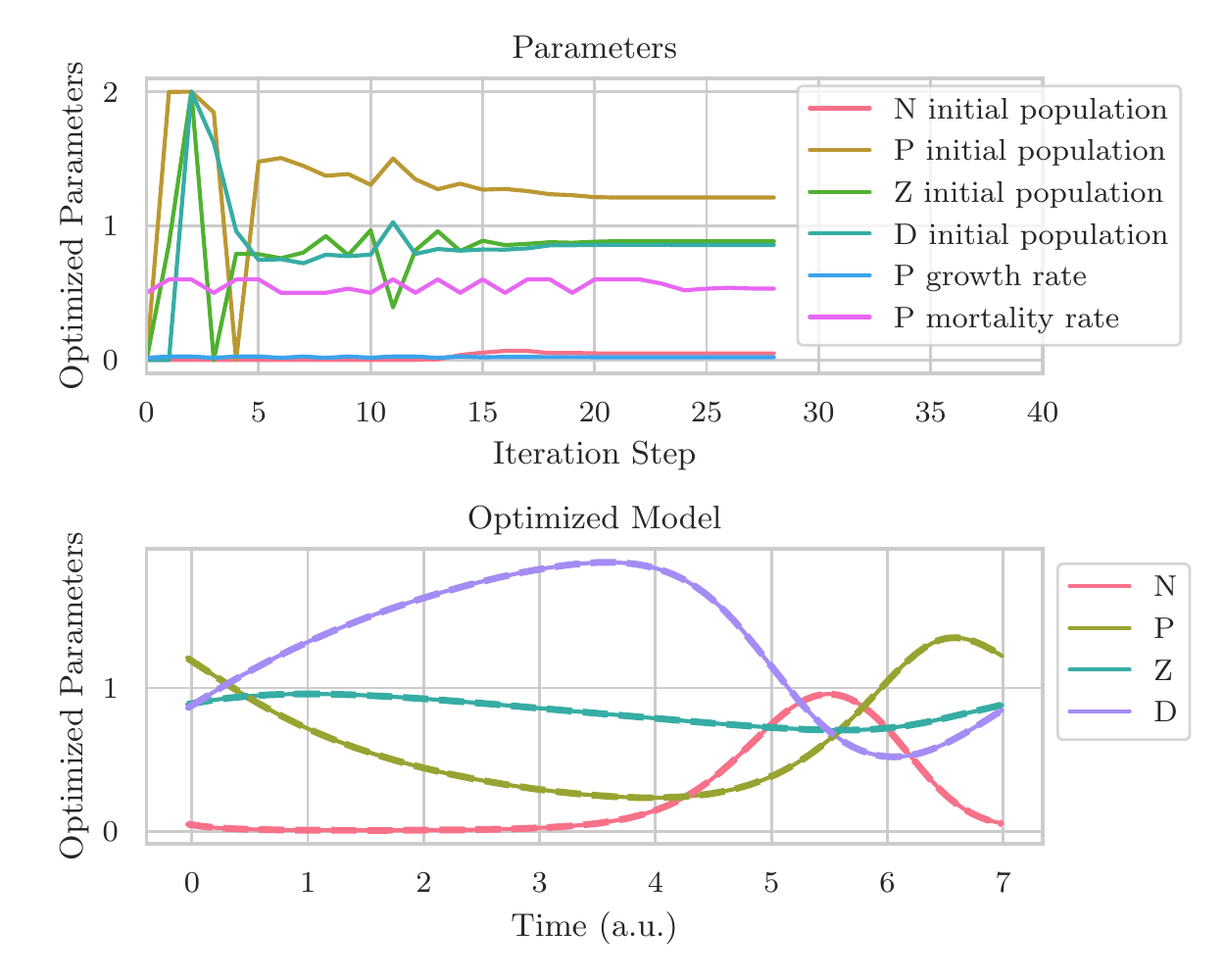}
  \caption{Exemplary output of a typical inverse modelling run. The lower figure shows the fitted model with dashed lines representing the reference data. The upper figure shows the model parameter configurations tested during the fitting process.}
  \label{fig:examplary_results}
\end{figure}

\paragraph{} 
After the framework has been configured and run, there are two possible kinds of output depending on the configuration.
In the case where no optimization was applied, the framework returns the time evolution of the model.
In the second case, when the model was fitted to data, the time integration of each iteration step, its associated parameters which configure the model, and costs
are returned.
In both cases, a plot is automatically generated to visualize the output.
For the  exemplary case, where we fit the NPZD model, the plot is presented in fig. \ref{fig:examplary_results}.
The output can also be directly passed on for further processing, e.g. a 
stability analysis.

\section*{Discussion}

The combination of the three steps:
\begin{itemize}
  \item conceptual design through a network
  \item numerical integration of the underlying non-linear differential equations
  \item inverse modelling
\end{itemize}
is a novel approach to ecosystem modelling which has not been used in any 
ecosystem modelling framework that is not tailored to a very specific application,
i.e. fishery management.
In the following we will discuss the benefits and drawbacks of this method.

\paragraph{}

As discussed above, a major benefit of this framework is time that is saved when
using it compared to writing a model yourself. The configuration of the framework 
for a large model, including its defining fitting constraints, can be done in 
mere minutes. 
Through that the time that is needed from initially drawing up a model,
to the point where the results can be analyzed is reduced to less then an hour. 
Writing a comparable implementation from scratch is expected to take several 
days to a week. 
Therefore, this framework is ideally suited for testing different 
implementations of nonlinear model interactions.
  
Due to its open design it is well suitable for a wide range of applications, 
including management activities. 
This is not limited to fishery management as many other frameworks 
but can also be applied in the area of e.g. blue carbon management.

Because of its simple design is also well suitable for teaching.
It can be used with limited understanding of the inner workings. 
Hence, this allows it to be used without in depth programming knowledge 
or understanding of the numerics applied and therefore assisting in 
leading students to the field of ecosystem modelling.

\paragraph{}
To this date, many ecosystem models and modelling frameworks used approximated,
linearized interactions to describe the studied ecosystem 
\cite{Melbourne-Thomas2012,Borrett2013,Borrett2014,Trebilco2020,Niquil2012}.
This is especially common for ecosystem-management applications.
The advantage of our approach is that it drastically simplifies the models and therefore makes them much easier to use and interpret the results.

However, many crucial interactions such as predator-prey interactions are non-linear,
as already shown in the famous Lotka-Volterra equations.
Hence, linearized models are crude approximations that are valid if, and only if,
used to study systems with only small changes in the compartment quantities.
Therefore, a non-linear description of a system is necessary if its full dynamics are studied.

\paragraph{} 
The NEMF framework offers both forward and inverse modelling methods.
However, an inverse modelling framework does not necessarily need to offer an 
option to solve the forward model.
It can be designed in such a way, that the user can provide an external 
forward model,
for example, the \textit{calibrar}\cite{Oliveros2016} framework. 
However, in that case the user needs to handle the communication between the 
forward model that passes the cost function for a certain model configuration to
the inverse modelling framework.
Therefore, they are required to write the parser code themselves.
This approach is sometimes referred to as \emph{blackbox} or \emph{out-of-box} 
approach.

\begin{figure}
  \includegraphics[width=\columnwidth]{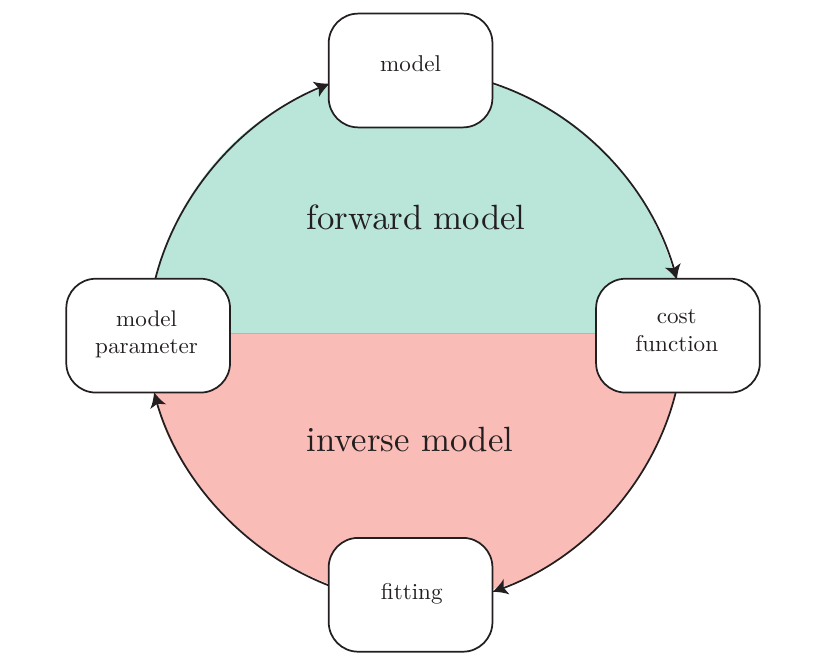}
  \caption{Schematic representation of an inverse modelling framework.
           If the upper forward modelling has to be provided by the user we call
           if a blackbox approach while if it is included inf the framework we 
           call it an inside-the-box approach. NEMF is the latter.}
  \label{fig:in-out-of-box}
\end{figure}

In our framework, the forward modelling is included.
This creates some restriction concerning the model because the user is not free to provide their own model anymore.
However, it offers the large benefit that the user does not \textit{need} to 
write any code and does not \textit{need} to understand how the inverse framework is implemented, 
because they do not need to write any parsers.
We refer to this as \emph{inside-the-box} approach (See fig. \ref{fig:in-out-of-box}).

\paragraph{}

While python offers many benefits, both for developing as well as for using this
framework, its computational speed is significantly slower compared to e.g. C or
FORTRAN.
However, because we limit ourself to box-models this is not an issue as 
computational times even for large single model runs is quasi-instantaneous. 
Typically, full optimization runs with NEMF are performed in time scales of
seconds to several minutes for larger models.


The framework is designed for deterministic models exclusively. 
Therefore, stochastic processes are not supported.
However, we expect it to be possible for the user to implement stochastic 
processes if needed. 
Yet, during the development we do not test stochastic functions.
Hence, we do not guarantee that they work as expected.


\paragraph{}
While we only discussed the framework in the context of ecosystem modelling 
applications, we want to mention that any kind of ODE system can be modelled in 
the same way.
We illustrated this in the framework documentation with the example of an 
enzymatic reaction and coupled oscillated masses.

\section*{Software Availability}
The framework is published as open-source software under the BSD-3-Clause License.
The source code can be found on \url{github.com/465b/nemf}
It can be acquired as a python package and listed in the Python Package Index (PyPI) 
under the acronym \textit{NEMF}.
For more information, see the official docs at \url{nemf.readthedocs.io}.

\section*{Acknowledgement}

We want to thank Maike Scheffold for inspiring this work and offering her feedback during the development.
We also want to thank Jan Byd{\v z}ovsk{\' y} and Klaus Gy{\"o}rgyfalvay for all the 
discussions that helped further this  project.